\documentclass[11pt]{article}
\usepackage{xcolor}
\usepackage{cite}
\usepackage{multirow}
\usepackage{enumerate}
\usepackage[scale=0.8]{geometry}

\usepackage{graphicx}
\usepackage[subrefformat=parens,labelformat=parens]{subfig}
\usepackage{hyperref}

\usepackage{amsfonts,amsmath}
\usepackage{multirow,multicol}
\usepackage{bbm}
\usepackage{xspace}
\usepackage{slashed}

\usepackage[sort&compress, english]{cleveref}

\newcommand{\cA}{\mathcal{A}}

\numberwithin{equation}{section}

\begin{document}
\vspace{1cm}
\begin{center}
{
\LARGE {\bf Approximate Ricci-flat Metrics for Calabi-Yau Manifolds}\\[12pt]
\vspace{1cm}
\normalsize
{\bf{Seung-Joo Lee$^{a,}$}\footnote{seungjoolee@yonsei.ac.kr}}
{\bf and}
{\bf{Andre Lukas$^{b,}$}\footnote{andre.lukas@physics.ox.ac.uk}}
\bigskip}\\[0pt]
\vspace{0.23cm}
${}^a$ {\it 
Department of Physics, Yonsei University, Seoul 03722, Republic of Korea
}\\[2ex]
${}^b$ {\it 
Rudolf Peierls Centre for Theoretical Physics, University of Oxford\\
Parks Road, Oxford OX1 3PU, UK
}\\[2ex]
\end{center}
\vspace{0.5cm}

\begin{abstract}\noindent
We outline a method to determine analytic K\"ahler potentials with associated approximately Ricci-flat K\"ahler metrics on Calabi-Yau manifolds. Key ingredients are numerically calculating Ricci-flat K\"ahler potentials via machine learning techniques and fitting the numerical results to Donaldson's Ansatz. We apply this method to the Dwork family of quintic hypersurfaces in $\mathbb{P}^4$ and an analogous one-parameter family of bi-cubic CY hypersurfaces in $\mathbb{P}^2\times\mathbb{P}^2$. In each case, a relatively simple analytic expression is obtained for the approximately Ricci-flat K\"ahler potentials, including the explicit dependence on the complex structure parameter. We find that these K\"ahler potentials only depend on the modulus of the complex structure parameter.
\end{abstract}

\setcounter{footnote}{0}
\setcounter{tocdepth}{2}
\clearpage
\tableofcontents

%%%%%%%%%%%%%%%%%%%%%%%%%%%%%%%%%%%%%%%%%%%%%%%%%%%%%%%%%%%%%%%%%%%%%%%%%%%%
\section{Introduction}\label{sec:introduction}
Yau's theorem guarantees the existence of a unique Ricci-flat K\"ahler metric with a given K\"ahler class on a Calabi-Yau (CY) manifold. Such Ricci-flat metrics are of mathematical interest and they play an important role in compactifications of string theory. Unfortunately, for compact Calabi-Yau manifolds of complex dimension three or higher, analytic expressions for Ricci-flat metrics are not known. Over the past few years substantial progress has nevertheless been made in numerically computing Ricci-flat metrics on Calabi-Yau three-folds, starting with Donaldson's algorithm~\cite{donaldson} and its applications~\cite{Douglas:2006hz,Douglas:2006rr,Braun:2007sn,Braun:2008jp,Anderson:2010ke,Anderson:2011ed,Headrick:2009jz,Cui:2019uhy} and, more recently, using machine learning methods~\cite{Ashmore:2019wzb,Douglas:2020hpv,Anderson:2020hux,Jejjala:2020wcc,Larfors:2021pbb,Ashmore:2021ohf,Larfors:2022nep}.

The Ricci-flat metric in numerical form is already useful, enabling us to compute the spectrum of the Laplacian on a CY manifold~\cite{Braun:2008jp,Ruehle:2024ufw} or the masses of quarks in a CY string compactification~\cite{Constantin:2024yxh}, to name just two applications. However, it would be all the more helpful and exciting to deal with the metric analytically. The purpose of this paper is to present analytic expressions for metrics (or, rather, their associated K\"ahler potentials) on relatively simple CY manifolds, which are approximately Ricci-flat. More specifically, we will find such metrics for the standard one-parameter Dwork family of quintics, as well as an analogous one-parameter family for the bi-cubic CY. For these cases, we will obtain a family of approximately Ricci-flat analytic metrics as an explicit function of the single complex structure modulus. 

The starting point of our discussion is Donaldson's Ansatz for CY metrics. Consider a CY manifold $X$, an ample line bundle $L\rightarrow X$ and a basis $\{s^I\}$ of sections of $\Gamma(X,L^k)$, where $k$ is a positive integer. Then, Donaldson's Ansatz for the K\"ahler potential $K$ reads
\begin{equation}\label{Kans0}
 K=\frac{1}{\pi k}\log(\kappa)\;,\qquad \kappa=H_{I\bar{J}}s^I\bar{s}^{\bar{J}}\; , 
\end{equation}
where $H$ is a Hermitian matrix to be determined. The associated K\"ahler form and the metric are given by $J=i\partial\bar{\partial}K$ and $g_{\alpha\bar{\beta}}=\partial_\alpha\partial_{\bar{\beta}}K$, respectively. Donaldson's algorithm involves finding, for any fixed $k$, a `balanced' metric, using an iterative procedure. It can be shown that these balanced metrics converge to the Ricci-flat metric with K\"ahler class $[J]$ proportional to $c_1(L)$, as $k\rightarrow\infty$. Practical numerical calculations can only be carried out up to a finite $k$ (as the dimension of $\Gamma(X,L^k)$ increases) and reasonable numerical approximations to the Ricci-flat metric by balanced metrics can be obtained for moderately large $k$, say $k\geq 6$~\cite{Douglas:2006rr}. Here, we will not attempt to find the balanced metric but ask instead about the `optimal' metric for a given $k$, that is, the metric within the family~\eqref{Kans0} which is `closest' to the Ricci-flat one. As we will see, this already leads to fairly accurate results for relatively small $k$, specifically $k=1,2,3$.

In practice, we proceed as follows. Our chosen CY manifolds, the one-parameter families of quintics and bi-cubics, have relatively large discrete symmetries $G$. Restrictions thus arise on the form of the Hermitian matrix $H$ in the K\"ahler potential~\eqref{Kans0}, which we first clarify. Next, we use machine learning techniques, specifically the {\sf cymetric} package~\cite{Larfors:2021pbb}, to compute the K\"ahler potential of the Ricci-flat metric numerically, for many choices of the single complex structure parameter $\psi$. We then fit the Ansatz~\eqref{Kans0} to each of these numerical K\"ahler potentials and thereby determine the remaining coefficients (after having imposed the symmetry constraints mentioned above) of $H$ as a function of $\psi$. In this way, we obtain remarkably accurate fits even for $k=2$ and a relatively simple K\"ahler potential of the form~\eqref{Kans0} with explicit $\psi$-dependence, leading to the associated analytic expression for an approximately Ricci-flat metric.

The plan of the paper is as follows. In the next section we introduce our method in detail by illustrating it with the standard one-parameter Dwork family of quintics. In Section~\ref{sec:bicubic} we consider a somewhat more complicated example, a one-parameter family of bi-cubic CY manifolds. We conclude in Section~\ref{sec:conclusion}.

\section{Approximate Ricci-flat metrics on the quintic}\label{sec:quintic}
Our strategy starts from choosing a one-parameter family of quintics $X$ with a symmetry group $G$ and restricting the K\"ahler potential Ansatz~\eqref{Kans0} by imposing $G$-invariance. This will reduce the number of free parameters in the Ansatz considerably and simplify the numerical calculation. 

In fact the procedure here is not quite as innocent as it might seem, because $G$ may in general have fixed points. 
If $G$ is freely-acting the quotient $X/G$ is a smooth CY manifold with a unique Ricci-flat metric of its own, which, pulled back, gives rise to a Ricci-flat metric on $X$. In this case, the associated Ricci-flat K\"ahler potential should thus be $G$-invariant by construction. However, the symmetry $G$ under consideration for our family of quintics is not freely-acting and, hence, the simple argument above does not apply to the current case. Nevertheless, it is reasonable to expect $G$-invariance of the K\"ahler potential even in this general situation and this is what we will assume. The success of our numerical computations serves as an a posteriori justification for this assumption. 

For the explicit computation, we first obtain approximately Ricci-flat metrics (and the associated K\"ahler potentials) numerically, for many values of the complex structure parameter $\psi$, using the {\sf cymetric} package. The results for any given $\psi$ are then fit to the K\"ahler potential Ansatz~\eqref{Kans0} (in its $G$-invariant form). Combining the results for all values of $\psi$, we finally determine the $\psi$-dependence of the K\"ahler potential.

\subsection{Quintic three-folds}
We begin by introducing the properties of quintic CY three-folds, relevant to our discussion; see, for example, Ref.~\cite{huebsch} for more information on the quintic, as well as other related Calabi-Yau manifolds, and Ref.~\cite{joyce} for mathematical background on manifolds with special holonomy.
Quintic CY three-folds $X$ have non-trivial Hodge numbers $(h^{1,1}(X),h^{2,1}(X))=(1,101)$ and Euler number $\eta(X)=-200$. They can be defined by the vanishing locus of a quintic polynomial in complex projective space $\mathbb{P}^4$, on which we introduce homogeneous coordinates $x_A$ with $A=0, \ldots, 4$. On a local patch, for example the one defined by $x_0\neq 0$, we have affine coordinates $x_a/x_0$ with $a=1,\ldots , 4$. For our purposes, we consider the one-parameter Dwork family of quintics, whose defining polynomials are given by
\begin{equation}\label{P}
 P=x_0^5+x_1^5+x_2^5+x_3^5+x_4^5-5\psi\, x_0x_1x_2x_3x_4\;, 
\end{equation}
where $\psi\in\mathbb{C}$ is the single complex structure parameter (out of a total of $h^{2,1}(X)=101$). 

Note that we may impose the following identification on the complex structure, 
\begin{equation} 
\psi\sim \alpha\psi\,, \quad \text{for}~\alpha=\exp(2\pi i/5)\,, 
\end{equation} 
since the coordinate transformation $x_0\mapsto \alpha x_0$, $x_a\mapsto x_a$ for $a=1,\ldots ,4$, combined with $\psi\mapsto \alpha^{-1}\psi$, leaves the polynomial~\eqref{P} invariant. In practice, this means we can restrict our calculation to the fundamental domain $0\leq{\rm arg}(\psi)<2\pi/5$. Within this domain the quintic~\eqref{P} defines a smooth manifold, except at the conifold point $\psi=1$. 

The above family of quintics has the well-known freely-acting $\mathbb{Z}_5^{(1)}\times \mathbb{Z}_5^{(2)}$ symmetry (projectively) generated by
\begin{equation}
 \mathbb{Z}_5^{(1)}\; :\; x_A\mapsto \alpha^A x_A
 \;,\qquad
 \mathbb{Z}_5^{(2)}\; :\; x_A\mapsto x_{(A+1)\,{\rm mod}\,5}\; .
 \end{equation}  
In fact, there is an even larger, non-freely acting symmetry group, 
\begin{equation}
G=\langle \mathbb{Z}_5^{(1)},S_5\rangle\,,   
\end{equation}
where $S_5$ permutes the homogeneous coordinates $x_A$, which leaves the polynomials~\eqref{P} invariant.

\subsection{K\"ahler potential Ansatz}
Choosing $L={\cal O}_X(1)$ and starting with $k=1$ we should consider the sections in $\Gamma(X,{\cal O}_X(1))$, a five-dimensional space spanned by the coordinates $x_A$, where $A=0,\ldots ,4$. In this case, the Ansatz~\eqref{Kans0} reads
\begin{equation}\label{Kquintic1}
 K=\frac{1}{\pi }\ln\left(\sum_{A,\bar{B}=0}^4H_{A\bar{B}}x_A\bar{x}_{\bar{B}}\right)\; ,
\end{equation}
where $H$ is a Hermitian $5\times 5$ matrix. For quintics defined by the polynomials~\eqref{P} we expect $K$ to respect the symmetry $G$, so we should consider only the $G$-singlets in $\Gamma(X,{\cal O}_X(1))\times\Gamma(X,{\cal O}_X(1))^*$. There is evidently only one such singlet which corresponds to a matrix $H$ proportional to the unit matrix. After absorbing the factor of proportionality into a projective coordinate re-scaling the K\"ahler potential becomes
\begin{equation}\label{FS-1}
 K=\frac{1}{\pi }\ln\left(\sum_{A=0}^4|x_A|^2\right)\; ,
\end{equation}
that is, the standard Fubini-Study K\"ahler potential.\footnote{This also works for general quintics since the matrix $H$ in Eq.~\eqref{Kquintic1} can be diagonalised to a unit matrix using a suitable ${\rm GL}(5)$ transformation acting on the homogeneous coordinates.}  No degree of freedom of the matrix $H$ remains.
This Fubini-Study K\"ahler potential (restricted to the quintic, using $P=0$) represents a (rather crude) approximation to the Ricci-flat K\"ahler potential. Perhaps surprisingly, even this crude approximation leads to reasonable results for the purpose of some calculations~\cite{Constantin:2024yxh}. 

To proceed more systematically, we incorporate the dependence of $K$ on the complex structure parameter $\psi$ as well as the K\"ahler parameter $t$, by using the $k=1$ Ansatz
\begin{equation}\label{Kans1}
 K=c\, t\ln(\kappa )\;,\qquad \kappa = a_0 I_0\,,
\end{equation} 
where $I_0$ denotes the $G$-singlet, 
\begin{equation}
I_0=\sum_{A=0}^4|x_A|^2\;,  
\end{equation}
and the parameter $c \in \mathbb R$ 
accounts for the overall normalisation. The single remaining degree of freedom of the matrix $H$, the parameter $a_0=a_0(\psi)$, amounts to a K\"ahler transformation. As a result, the metric associated to Eq.~\eqref{Kans1} is $\psi$-independent, as expected from the above arguments.

For a better approximation of the Ricci-flat metric, we proceed naturally to $k=2$ and sections of $\Gamma(X,{\cal O}_X(2))$, a $15$-dimensional space spanned by the quadratic monomials $x_Ax_B$, where $A,B=0,\ldots , 4$ and $A\leq B$. Again, we expect $K$ to respect the symmetry $G$ and focus on the $G$-singlets in $\Gamma(X,{\cal O}_X(2))\times\Gamma(X,{\cal O}_X(2))^*$.  As it turns out, there are precisely two such singlets, namely
\begin{equation}\label{Iquintic2}
I_0=\sum_{A<B}|x_A|^2|x_B|^2\; ,\qquad   I_1=\sum_A|x_A|^4\,.
\end{equation}  
Therefore, the Ansatz~\eqref{Kans0} for the K\"ahler potential can be written as
\begin{equation}\label{Kans2}
 K=c\,t\ln(\kappa)\;,\qquad \kappa=a_0 I_0+a_1 I_1
\end{equation}
where, again, $t$ is the single K\"ahler parameter and $c\in\mathbb{R}$ a constant, while the two remaining degrees of freedom, $a_0=a_0(\psi)$ and $a_1=a_1(\psi)$ of the matrix $H$, are real functions of the complex structure parameter $\psi$. Our goal is to determine the functional forms of $a_0(\psi)$ and $a_1(\psi)$ by a comparison with numerical calculations.

Before we do so, let us also discuss the situation at the next higher order, $k=3$. In this case, we should consider sections of $\Gamma(X,{\cal O}_X(3))$, a $35$-dimensional space spanned by the cubic monomials $x_Ax_Bx_C$, where $A,B,C=0,\ldots ,4$ and $A\leq B\leq C$. The space $\Gamma(X,{\cal O}_X(3))\times \Gamma(X,{\cal O}_X(3))^*$ contains three $G$-invariants which are given by
\begin{equation}\label{Iquintic3}
 I_0=\sum_{A<B<C}|x_A|^2|x_B|^2|x_C|^2\;,\qquad
 I_1=\sum_{A\neq B}|x_A|^4|x_B|^2\;,\qquad
 I_2=\sum_A|x_A|^6 \; .
\end{equation} 
This means that the Ansatz for the Kahler potential is now of the form
\begin{equation}\label{Kans3}
 K=c\,t\ln(\kappa)\;,\qquad \kappa=a_0 I_0+a_1 I_1+a_2 I_2\; ,
\end{equation}
where, as before, the coefficients $a_r=a_r(\psi)$ with $r=0,1,2$ are real functions of the complex structure parameter $\psi$.

\subsection{Numerical results}
We would like to consider quintics defined by Eq.~\eqref{P} for values of the complex structure parameter $\psi=0.4\, n \exp(2\pi i m/25)$, where $n=0,\ldots 50$ and $m=0,\ldots  5$, leading to a total of $301$ points and covering the fundamental domain\footnote{Strictly speaking, any values of the complex structure $\psi$ with $m=5$ do not lie in the fundamental domain and should rather be considered redundant, as they are identified with their $m=0$ counterparts. Nevertheless, we still include them in our analysis for a sanity check.} $0\leq{\rm arg}(\psi)<2\pi/5$ in the range $0\leq |\psi|\leq 20$. The single K\"ahler parameter is set to $t=1$; the K\"ahler potential and the metric scale with $t$, so the $t$-dependence can easily be restored later. For each of the above $\psi$-values, we use the {\sf cymetric} point generator to generate a data set of $100000$ points $p_i$ on the manifold. We have checked that these data sets respect the $G$ symmetry to good accuracy, that is, for each data point $p$ the data set also contains points close to the orbit points $Gp$.

\begin{figure}[h]
\begin{center}
\includegraphics[width=0.46\textwidth]{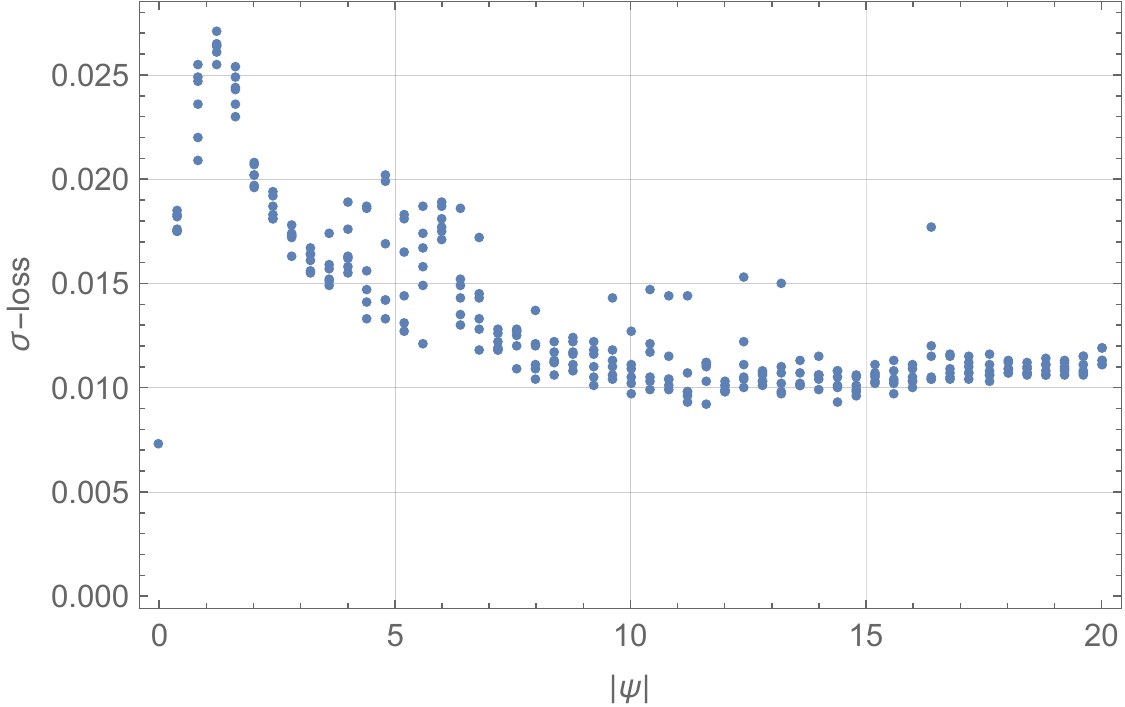}\hskip 10mm
\includegraphics[width=0.46\textwidth]{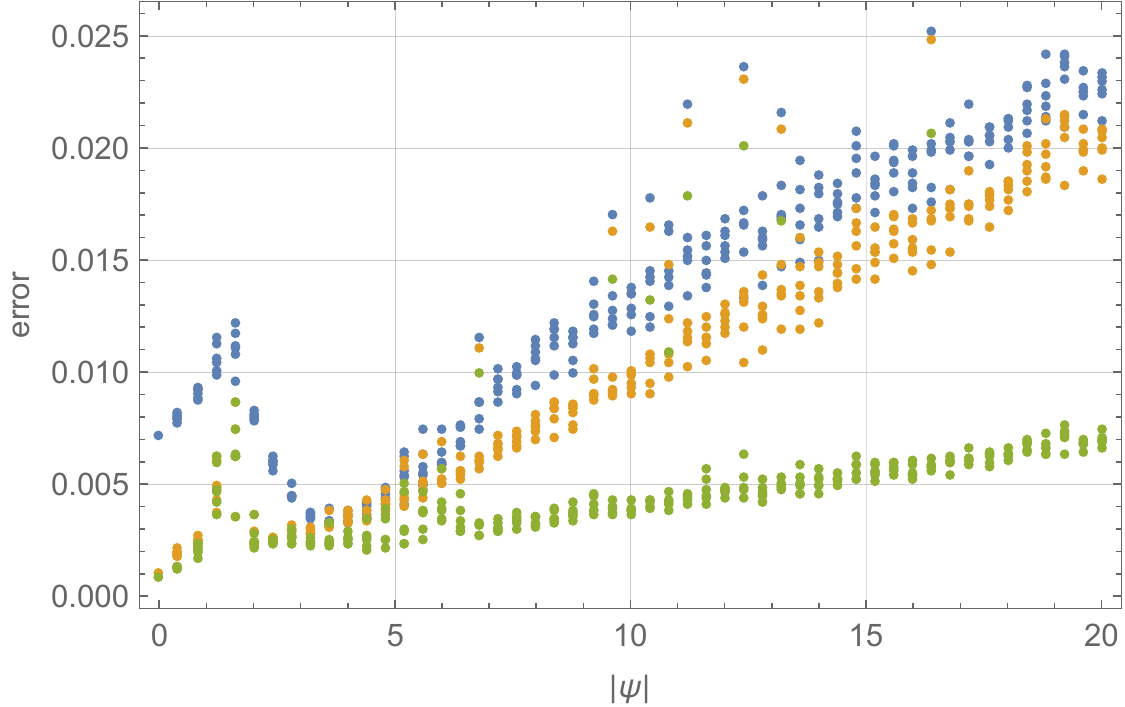}
\caption{Left plot: Final $\sigma$-loss as a function of $|\psi|$ for the quintic. Right plot: Mean of the relative error $|(K_i-K(p_i))/K(p_i)|$ as a function of $|\psi|$ for the K\"ahler $k=1$ potential~\eqref{Kans1}(blue), the $k=2$ K\"ahler potential~\eqref{Kans2} (yellow) and the $k=3$ K\"ahler potential~\eqref{Kans3} (green).}\label{fig:quinticsigmaloss}
\end{center}
\end{figure}

Using the {\sf cymetric} $\phi$-model we train a fully-connected neural network (width $64$, depth $3$, GeLU activation) for each of the $301$ data sets for $120$ epochs. The final Monge-Ampere loss, as defined in Ref.~\cite{Larfors:2021pbb}, as a function of $|\psi|$ is shown in Fig.~\ref{fig:quinticsigmaloss} on the left. It is in the range $0.01$ -- $0.025$, indicating that we have found Ricci-flat K\"ahler metrics to a good approximation. Using the trained networks, we read out the pairs $(p_i,K_i)$, where $K_i$ denote the values of the numerical K\"ahler potential evaluated at the points $p_i$. We may then use this K\"ahler potential data to determine the coefficients $c$ and $a_r$ in the Ans\"atze~\eqref{Kans1}, \eqref{Kans2} and \eqref{Kans3} as follows. 

We begin with the somewhat trivial case $k=1$, that is, the Fubini-Study metric~\eqref{Kans1}. We have performed a fit to this Ansatz for all $301$ data sets, each consisting of 100000 pairs $(p_i,K_i)$, to determine the optimal choices for the parameters $c$ and $a_0$ in each case. The value of $a_0$ is not meaningful as it corresponds to a K\"ahler transformation and that of $c$ is found to be approximately $\psi$-independent, $c\simeq 1/\pi$, in accordance with Eq.~\eqref{Kans0}. The mean of the relative deviation $|(K_i-K(p_i))/K(p_i)|$, evaluated from Eq.~\eqref{Kans1} with the best-fit values for $c$ and $a_0$ inserted, is shown in Fig.~\ref{fig:quinticsigmaloss} on the right (blue) for each $|\psi|$. 

We move on to the case $k=2$. As before, for all $301$ data sets $\{(p_i,K_i)\}$ we have performed a fit to the K\"ahler potential Ansatz~\eqref{Kans2}, thereby determining the optimal parameters $c$ and $a_i$ in each case. The error for this Ansatz with the best-fit values for the parameters, defined again as the mean of the relative deviation, is shown in Fig.~\ref{fig:quinticsigmaloss} on the right (yellow) for each $|\psi|$. This error is about 2\% at most for larger values of $|\psi|$ and somewhat smaller, around 1\%, for small $|\psi|$-values. This means, Donaldson's Ansatz for $k=2$ can already provide an approximation to the Ricci-flat K\"ahler potential at the percent level, at least for our quintic example.

\begin{figure}[h]
\begin{center}
\includegraphics[width=0.45\textwidth]{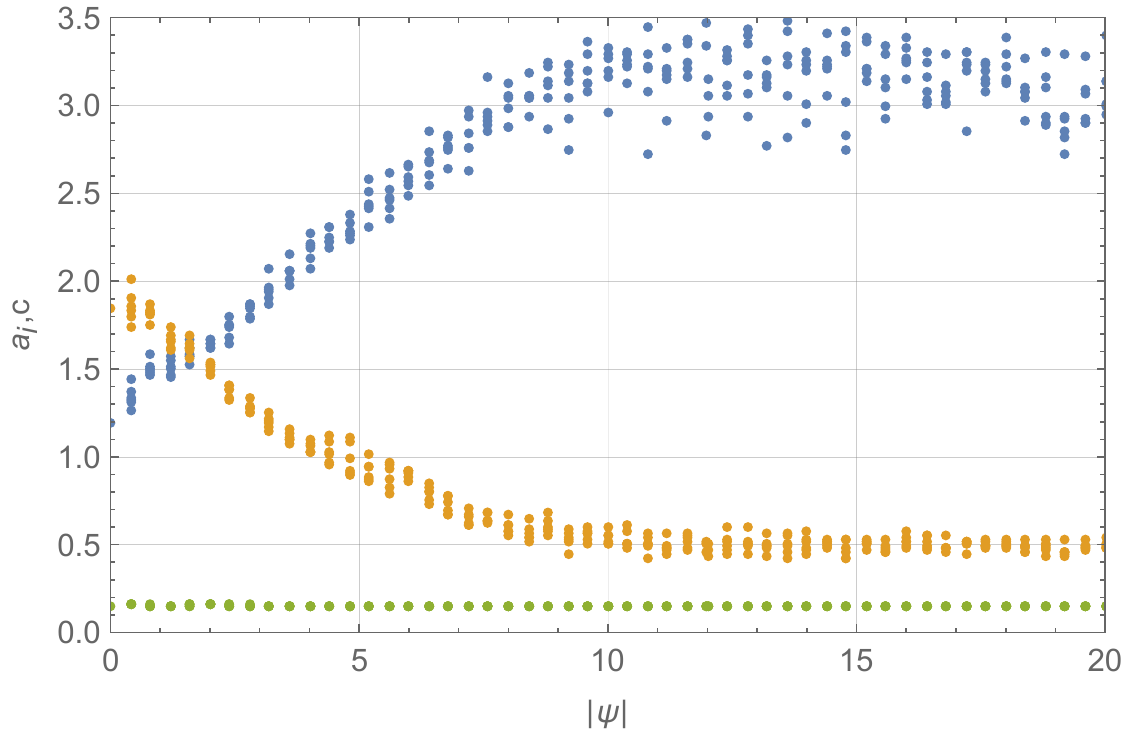}\hskip 10mm
\includegraphics[width=0.45\textwidth]{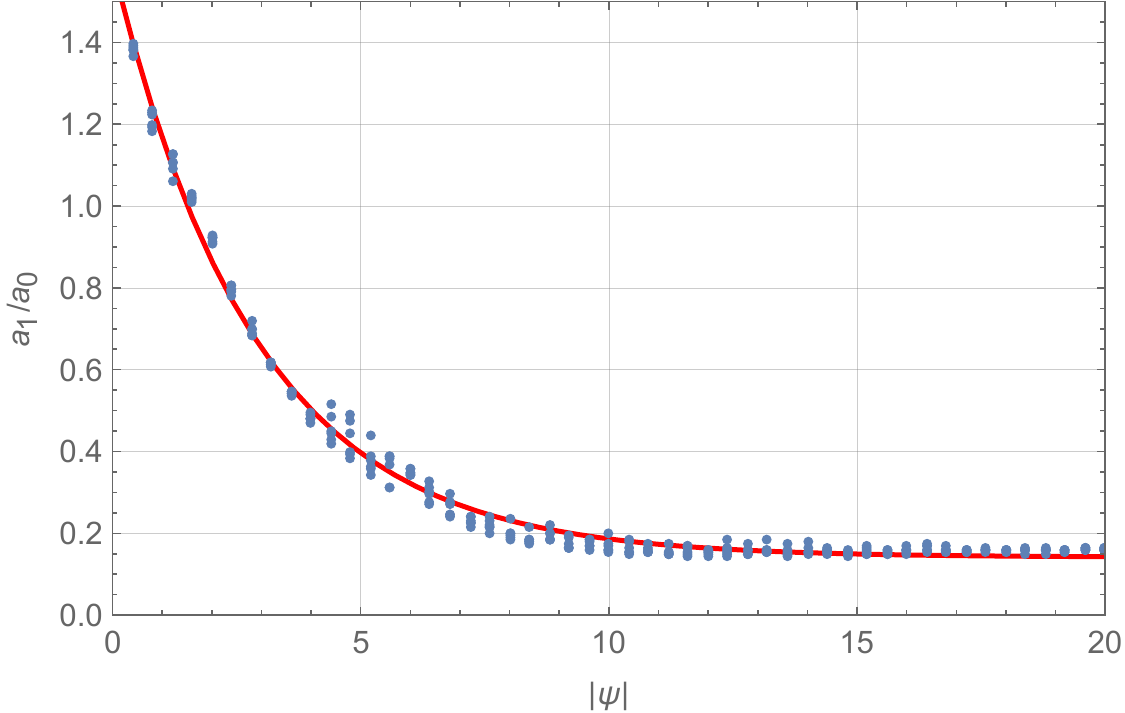}\hskip 10mm
\caption{Left plot: Best-fit coefficients $a_0(\psi)$ (blue) and $a_1(\psi)$ (orange) and $c$ (green) in the Ansatz~\eqref{Kans2} for the quintic K\"ahler potential, as a function of $|\psi|$. Right plot: Ratio $a_1(\psi)/a_0(\psi)$ as a function of $|\psi|$ and best-fit of the form~\eqref{a0a1fit}.}\label{fig:quinticapsi2}
\end{center}
\end{figure}

The optimal values for $c$ and $a_r$ as a function of $|\psi|$ for the $k=2$ Ansatz~\eqref{Kans2} are shown in Fig.~\ref{fig:quinticapsi2} on the left.
The first observation from this figure is that the coefficient $c$ is $\psi$-independent, as expected, and its value, $c\simeq \frac{1}{2\pi}$, is in good agreement with Eq.~\eqref{Kans0}. The other features in Fig.~\ref{fig:quinticapsi2} on the left look disappointing at first: the blue and the orange points which represent the optimal values of $a_0(\psi)$ and $a_1(\psi)$, respectively, have a rather substantial scatter, especially at larger values of $|\psi|$. There are two possible reasons why disappointment might be premature. First, there is no a-priori reason why the $a_i$ should be functions of the modulus $|\psi|$, rather than of the full complex variable $\psi$. Furthermore, the ambiguity of $K$ due to K\"ahler transformations implies that $a_0(\psi)$ and $a_1(\psi)$ do not carry independent meaning; rather, the K\"ahler potential~\eqref{Kans2} can, equivalently, be written as
\begin{equation}
 K=c\,t\ln\left(I_0+\frac{a_1(\psi)}{a_0(\psi)}I_1\right)+\mbox{K\"ahler transformation} \,, 
\end{equation}
which suggests plotting the ratio $\frac{a_1(\psi)}{a_0(\psi)}$ as opposed to the individual parameters $a_0(\psi)$ and $a_1(\psi)$. The resulting plot, as shown in Fig.~\ref{fig:quinticapsi2} on the right, strongly indicates that the ratio is a function of the modulus $|\psi|$ only. This function can, for example, be accurately fit by an exponential
\begin{equation}\label{a0a1fit}
 f(|\psi|)=\frac{a_1}{a_0}\simeq 0.141+1.453\, e^{-0.348\,|\psi|} \;,
\end{equation}
whose graph is given by the red curve in Fig.~\ref{fig:quinticapsi2}, on the right. This leads to the final form for the quintic K\"ahler potential,
\begin{equation}\label{kpq2}
 K=\frac{t}{2\pi}\ln\left(I_0+f(|\psi|)I_1\right)\; .
\end{equation}
Here, $I_0$ and $I_1$ have been defined in Eq.~\eqref{Iquintic2} and the $\psi$-dependence of $K$ is fully encoded in the functional behavior of $f$, for which we can use Eq.~\eqref{a0a1fit}. This is a rather simple expression and the metric associated to this K\"ahler potential is approximately Ricci-flat to within a few percent. 
 
\begin{figure}[h]
\begin{center}
\includegraphics[width=0.45\textwidth]{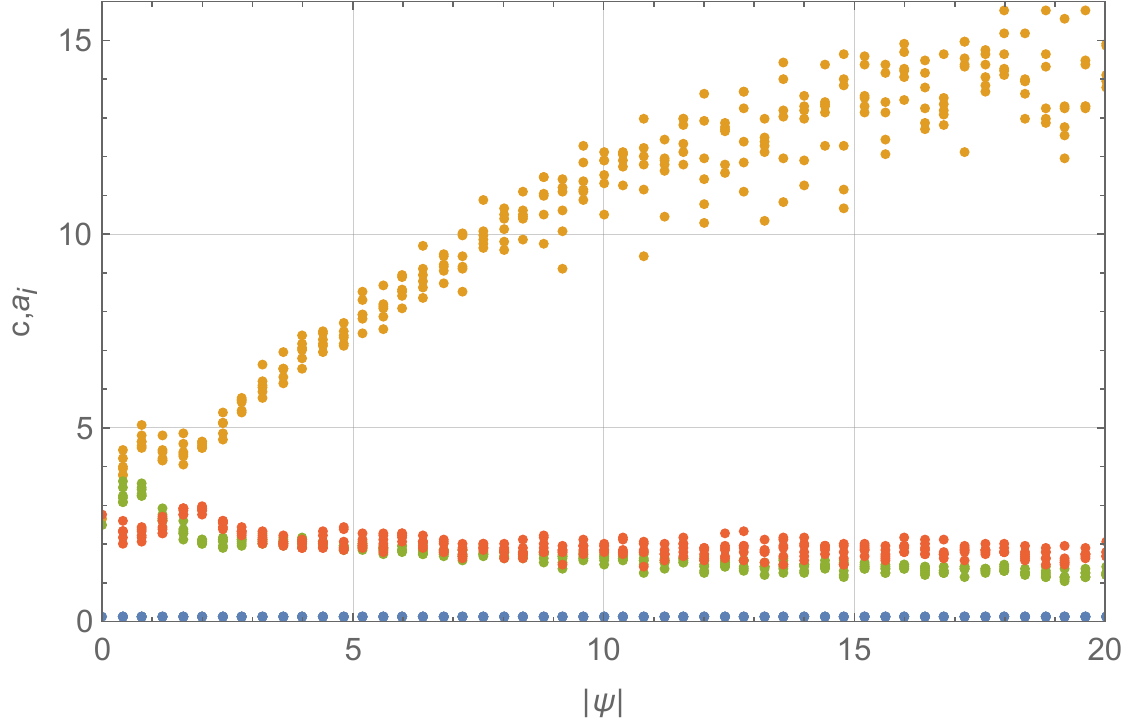}\hskip 10mm
\includegraphics[width=0.45\textwidth]{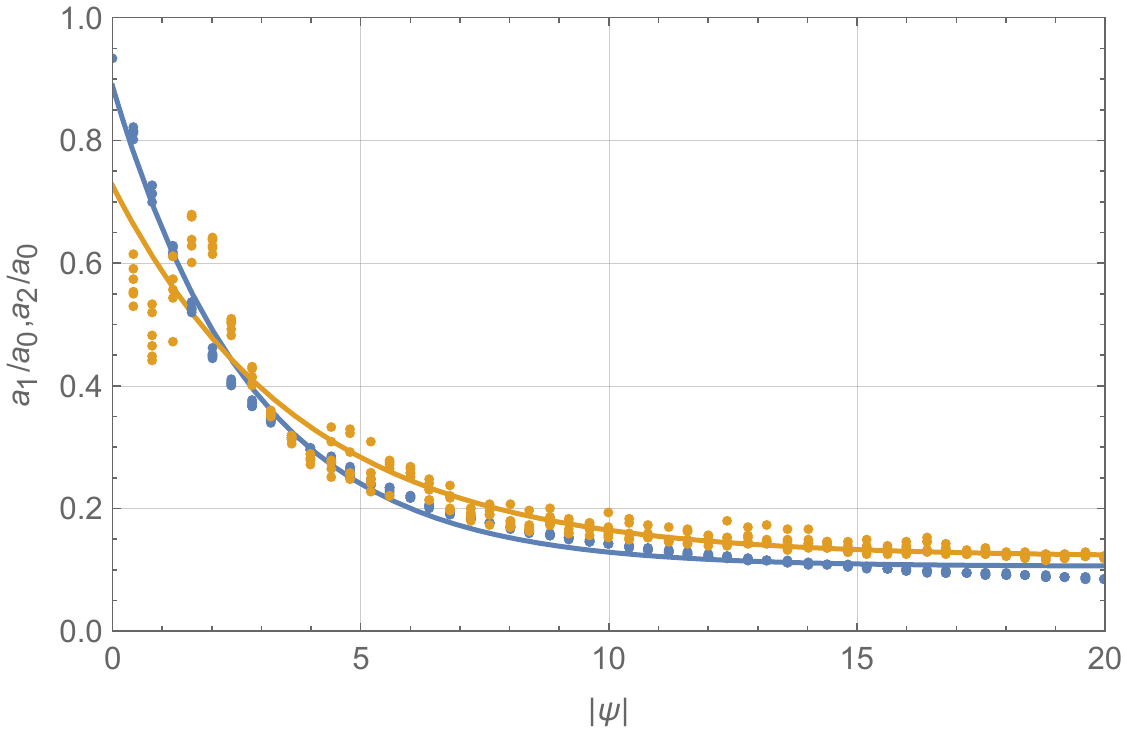}
\caption{Left plot: Best-fit coefficients $a_0(\psi)$ (yellow), $a_1(\psi)$ (red), $a_2(\psi)$ (green) and $c$ (blue) in the Ansatz~\eqref{Kans3} for the quintic K\"ahler potential, as a function of $|\psi|$. Right plot: The ratios $a_1(\psi)/a_0(\psi)$ (yellow points) and $a_2(\psi)/a_0(\psi)$ (blue points) as a function of $|\psi|$ and best fits of the form~\eqref{aia0}.}\label{fig:quinticapsi3}
\end{center}
\end{figure}

Let us extend this discussion to the case $k=3$, that is, to the K\"ahler potential Ansatz~\eqref{Kans3}, using the same data sets $\{(p_i,K_i)\}$ for complex structure parameters $\psi=0.4\, n \exp(2\pi i m/25)$, where $n=0,\ldots 50$ and $m=0,\ldots  5$, as before.
The best-fit values for $c$ and $a_r(\psi)$ as a function of $|\psi|$ are shown in Fig.~\ref{fig:quinticapsi3}. 
\noindent
The mean error between the data and the Ansatz~\eqref{Kans3} with these best-fit values inserted is shown in Fig.~\ref{fig:quinticsigmaloss} on the right (green). We note that this error is typically smaller than the one for $k=2$ (yellow) and is below $1\%$ for most $\psi$-values.
In conclusion, the Ansatz~\eqref{Kans3} provides a rather good approximation to the Ricci-flat K\"ahler potential for the quintic. 

From Fig.~\ref{fig:quinticapsi3} on the left, it is evident that the value of $c$ is $\psi$-independent, as expected, and is given by $c\simeq \frac{1}{3\pi}$, in accordance with Eq.~\eqref{Kans0}. Furthermore, the scatter in Fig.~\ref{fig:quinticapsi3} on the left can be reduced significantly by  considering the ratios $a_1(\psi)/a_0(\psi)$ and $a_2(\psi)/a_0(\psi)$, as  Fig.~\ref{fig:quinticapsi3} on the right shows. This provides good evidence that $a_1/a_0$ and $a_2/a_0$ are functions of $|\psi|$ only and it turns out that the functional dependence is well described by the exponentials
\begin{equation}\label{aia0}
f_1(|\psi|)=\frac{a_1}{a_0}\simeq 0.121+0.605\, e^{-0.264\,|\psi|} \; ,\qquad
f_2(|\psi|)=\frac{a_2}{a_0}\simeq 0.106+0.783\, e^{-0.352\, |\psi|}\; ,
\end{equation}
whose graphs are shown in Fig.~\ref{fig:quinticapsi3}, on the right.
Hence, the final form of the approximately Ricci-flat K\"ahler potential for the quintic is given by
\begin{equation}\label{kpq3}
 K=\frac{t}{2\pi}\log\left(I_0+f_1(|\psi|)I_1+f_2(|\psi|)I_2\right)\; ,
\end{equation}
with the invariants $I_r$ from Eq.~\eqref{Iquintic3} and the functions $f_1$, $f_2$ from Eq.~\eqref{aia0}.

\begin{figure}[h]
\begin{center}
\includegraphics[width=0.5\textwidth]{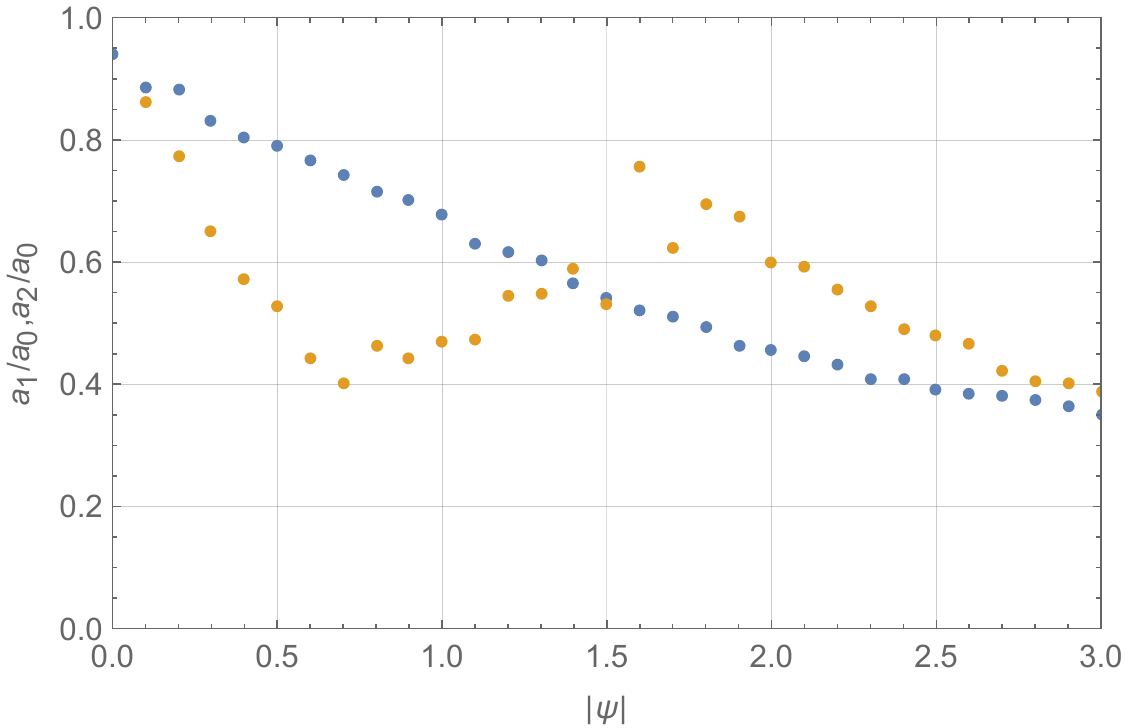}
\caption{The ratios $a_1(\psi)/a_0(\psi)$ (yellow) and $a_2(\psi)/a_0(\psi)$ (blue) for the quintic, as a function of $|\psi|$, focusing on values $\psi=0.1\,n$, for $n=0,\dots ,30$, near the conifold point $\psi=1$.}\label{fig:quinticsmallpsi}
\end{center}
\end{figure}

Let us note that Fig.~\ref{fig:quinticapsi3} on the right shows a somewhat irregular behaviour of $a_1(\psi)/a_0(\psi)$ for small values $0\leq|\psi|\leq 3$. We have thus looked at this feature in more detail by computing the numerical Ricci-flat metric for $\psi=0.1\, n$, with $n=0,\ldots ,30$ and fitting the results, as before, to the $k=3$ Ansatz~\eqref{Kans3}. The results for $a_1(\psi)/a_0(\psi)$ and $a_2(\psi)/a_0(\psi)$ in this small-$|\psi|$ range are shown in Fig.~\ref{fig:quinticsmallpsi},
\noindent
which indicate that $a_1(\psi)/a_0(\psi)$ does behave non-trivially around $\psi=1$. (This non-trivial feature is of course not captured by our simple fit in Eq.~\eqref{aia0}.)

It is tempting to attribute this behaviour to the conifold singularity the quintic~\eqref{P} exhibits at $\psi=1$. This may indeed be the case but, in view of our numerical approach, one has to be careful interpreting the effect of a singularity. For a choice of $\psi$ near $\psi=1$ a significant portion of the quintic's curvature is concentrated on a small locus near the (developing) conifold. As we approach $\psi=1$ this locus becomes smaller and, at some point, can no longer be resolved in our numerical calculation which relies on a finite point sample. In this case, the numerical metric is still a good approximate Ricci-flat metric in the `bulk' but it fails to describe the high-curvature conifold region. It is plausible that the non-trivial behaviour of $a_1(\psi)/a_0(\psi)$ around $\psi=1$ reflects this limitation of the numerical method to capture the small regions of high curvature. In summary, we expect our metric still provides an approximate Ricci-flat metric in the `bulk' but, near $\psi=1$, it does not account for the high-curvature locus around the conifold.
 
\section{Approximate Ricci-flat metrics on the bi-cubic}\label{sec:bicubic}
The basic strategy for bi-cubic CY manifolds follows the one for the quintic explained earlier.

\subsection{Bi-cubic three-folds}
We start with the ambient space $\cA=\mathbb{P}^2\times\mathbb{P}^2$. The ambient homogeneous coordinates are denoted by $x_A$ and $y_A$ with $A=0,1,2$, for the two $\mathbb{P}^2$ factors. On a local patch, for example the one where $x_0\neq 0$ and $y_0\neq 0$, we have the four affine coordinates $x_a/x_0$ and $y_a/y_0$ with $a=1,2$. A bi-cubic three-fold $X$ is then defined as the zero locus of a bi-degree $(3,3)$ polynomial in $\cA$. The non-trivial Hodge numbers are $(h^{1,1}(X), h^{2,1}(X))=(2,83)$, which imply the Euler number $\eta(X)=-162$. The K\"ahler class $[J]$ can be parametrized as $[J]=t_1[J_1]+t_2[J_2]$, where $J_1$ and $J_2$ are the restrictions of the Fubini-Study K\"ahler forms of the ambient $\mathbb{P}^2$ factors, and $t_1$, $t_2$ are the two K\"ahler parameters. In terms of these parameters, the K\"ahler cone is characterised by $t_1,t_2>0$.

There is a well-known family of bi-cubics with a freely-acting  symmetry defined by the cyclic generators, 
\begin{equation}
 \mathbb{Z}_3^{(1)}:\left\{\begin{array}{l}x_A\mapsto \alpha^A x_A\\
 y_A\mapsto \alpha^A y_A\end{array}\right.\;,\qquad
 \mathbb{Z}_3^{(2)}:\left\{\begin{array}{l}x_A\mapsto x_{(A+1)\,{\rm mod}\,3}\\
 y_A\mapsto y_{(A+1)\,{\rm mod}\,3}\end{array}\right. \; ,
\end{equation}
where $\alpha=\exp(2\pi i/3)$. If the above action is understood as a projective representation on $\mathbb{P}^2\times\mathbb{P}^2$ the two generators commute and the total group is $\mathbb{Z}_3^{(1)}\times \mathbb{Z}_3^{(2)}$. On the other hand, understood as a linear action on $\mathbb{C}^6$ the Schur cover of $\mathbb{Z}_3^{(1)}\times \mathbb{Z}_3^{(2)}$, a group of order $27$, is generated. 

For our applications, we are interested in a large, non-freely acting symmetry which contains the previous one as a sub-group. This symmetry contains the two cyclic groups generated respectively by
\begin{equation}\label{bicubicsymm}
 \mathbb{Z}_3^{(x)}:\left\{\begin{array}{l}x_A\mapsto \alpha^A x_A\\y_A\mapsto y_A\end{array}\right.\;,\qquad
 \mathbb{Z}_3^{(y)}:\left\{\begin{array}{l}x_A\mapsto  x_A\\y_A\mapsto \alpha^A y_A\end{array}\right.\;,
\end{equation}
as well as the symmetric group, $S_3^{(x,y)}$, consisting of  coordinate permutations of the form, 
\begin{equation}\label{bicubicsymm-S}
\left\{\begin{array}{l} x_A\mapsto x_{\sigma(A)}\\y_A\mapsto y_{\sigma(A)}\end{array}\right.\;,
\end{equation}
where $\sigma\in S_3$ is a permutation of $\{0,1,2\}$. Projectively, these generate a group of order $54$ isomorphic to $\mathbb{Z}_3^{(x)}\times\mathbb{Z}_3^{(y)}\times S_3^{(x,y)}$, while the linear action generates a group $G$ of order $486$. The sub-space of $G$-invariant bi-degree $(3,3)$ polynomials is five-dimensional and spanned by the following invariants.
\begin{equation}
\begin{array}{rclcrclcrcl}
 P_1&=&\sum_Ax_A^3y_A^3\;,&\quad&P_2&=&\sum_{A\neq B}x_A^3y_B^3\;,&\quad&&&\\[2mm] P_3&=&y_0y_1y_2\sum_A x_A^3\;,&&
 P_4&=&x_0x_1x_2\sum_Ay_A^3\;,&\quad&
 P_5&=&x_0x_1x_1y_0y_1y_2\; .
 \end{array}
\end{equation}
Within this space of $G$-invariants we focus on the one-parameter family of defining equations
\begin{equation}\label{bcp}
     P=P_1-P_2+3\psi\, P_3\;,
\end{equation}
somewhat analogous to the Dwork family of quintics, where $\psi\in\mathbb{C}$ is a single complex structure parameter. This family of bi-cubics is invariant under the coordinate transformation $x_A\mapsto x_A$, $y_0\mapsto\alpha y_0$, $y_a\mapsto y_a$, combined with $\psi\mapsto \alpha^{-1}\psi$. Therefore, the complex structure parameter should be identified as $\psi\sim\alpha\psi$. In practice, we restrict $\psi$ to the fundamental domain $0\leq{\rm arg}(\psi)<2\pi/3$.

\subsection{K\"ahler potential Ansatz}
For the bi-cubic the simplest choice is to start with the line bundle $L={\cal O}_X(1,1)$ and $k=1$. In particular, this means we are working on the line $t_1=t_2$ in K\"ahler moduli space. The space of sections $\Gamma(X,{\cal O}_X(1,1))$ is $9$-dimensional and spanned by the monomials $x_A\,y_B$, where $A,B=0,1,2$. There are only two $G$-singlets contained in $\Gamma(X,{\cal O}_X(1,1))\times \Gamma(X,{\cal O}_X(1,1))^*$ (where $G$ is the group generated by the actions in Eqs.~\eqref{bicubicsymm} and~\eqref{bicubicsymm-S}), namely
\begin{equation}\label{bcinv}
 I_0=\sum_{A=0}^2|x_A|^2|y_A|^2\;,\qquad
 I_1=\sum_{A\neq B}|x_A|^2|y_B|^2 \; .
\end{equation}
Hence, we can write the Ansatz for the K\"ahler potential as
\begin{equation}\label{bckp}
 K=c\,t\ln(\kappa)\;,\qquad\kappa=a_0I_0+a_1I_1\;,
\end{equation}
where $t$ is the K\"ahler parameter along the line $t_1=t_2$ in K\"ahler moduli space, that is, we take $t_1=t_2=t$.

\subsection{Numerical results}
We are considering the one-parameter family of bi-cubic three-folds defined by the zero locus of the polynomial $P$ in Eq.~\eqref{bcp}. In line with our Ansatz for the K\"ahler potential, we focus on K\"ahler parameters $t_1=t_2=1$ (with the single parameter $t$ along the line $t_1=t_2$ easily re-instated). For the complex structure parameter $\psi$ we consider values of the form $\psi=0.4 n\exp(2\pi i m/18)$, where $n=0,\ldots ,50$ and $m=0,\ldots ,5$, covering the fundamentail domain $0\leq {\rm arg}(\psi)<2\pi/3$ in the range $0\leq |\psi|\leq 20$, for a total of $301$ values. As before, we have used the {\sf cymetric} point generator to obtain $100000$ points $p_i$ on the manifold, for each $\psi$-value. 

\begin{figure}[h]
\begin{center}
\includegraphics[width=0.46\textwidth]{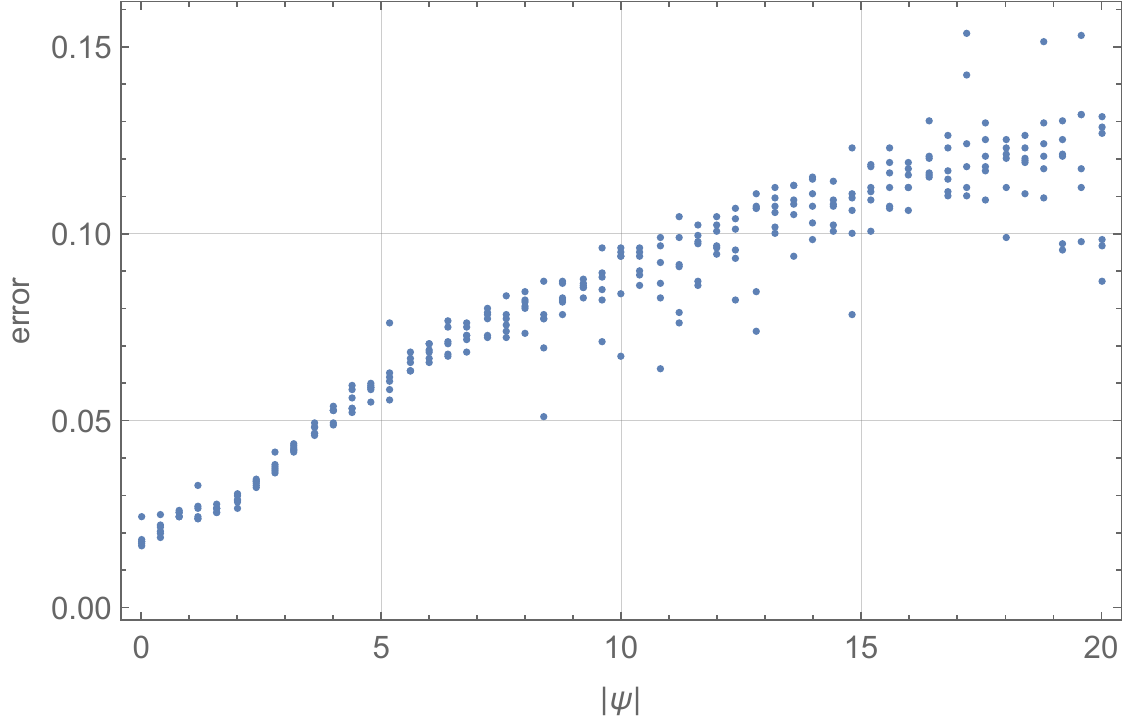}\hskip 10mm
\includegraphics[width=0.46\textwidth]{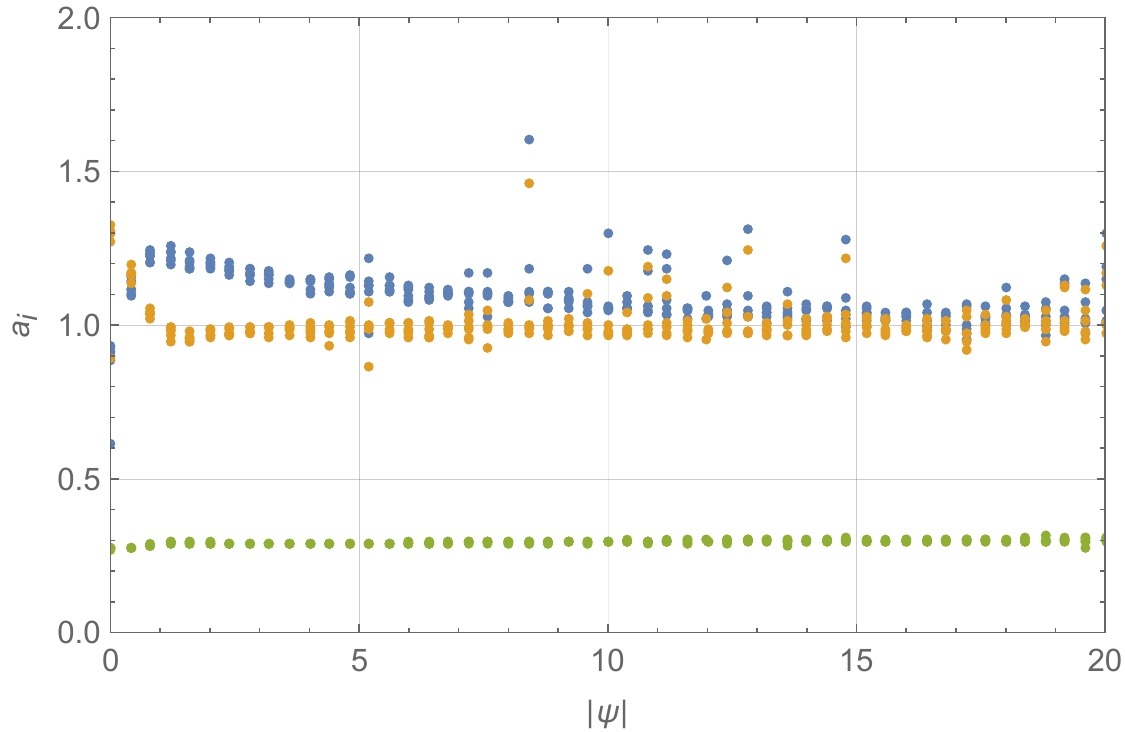}
\caption{Left plot: Mean of the relative deviation $|(K_i-K(p_i))/K(p_i)|$ as a function of $\psi$ for the best fit of the data $(p_i,K_i)$ to the bi-cubic Kahler potential Ansatz~\eqref{bckp}. Right plot: Best fit values of $a_0$ (blue), $a_1$ (yellow) and $c$ (green) in the Ansatz~\eqref{bckp}.}\label{fig:bcabc}
\end{center}
\end{figure}

With each of these data sets we have trained a {\sf cymetric} $\phi$-model (width $64$, depth $3$, GeLU activation) for $100$ epochs. The  pairs $(p_i,K_i)$,  $i=1,\ldots,100000$, of points $p_i$ and associated K\"ahler potential values $K_i$ have then been read out for each neural network and fitted to the Ansatz~\eqref{bckp}, from which we have found the values of the three parameters $a_0$, $a_1$ and $c$ for each $\psi$-value. The quality of the fit, measured by the mean of the relative deviation $|(K_i-K(p_i))/K(p_i)|$ (where the values $K(p_i)$ have been computed from Eq.~\eqref{bckp} with the best-fit values for $a_0$, $a_1$ and $c$ inserted), is shown as a function of $|\psi|$ in Fig.~\ref{fig:bcabc} on the left.
The error is relatively small, about a few percent, for small $\psi$ but grows with $|\psi|$ to around $10\%$. This increase correlates with an increase in final training loss for larger $|\psi|$ and is therefore related to the residual error in the trained neural networks. It is typical that training becomes more difficult for more extreme parameter choices. While the situation can in principle be improved by taking more point samples and/or performing more extensive training, we will not attempt to improve on this as the present results are sufficient for our purposes.

\begin{figure}[h!]
\begin{center}
\includegraphics[width=0.6\textwidth]{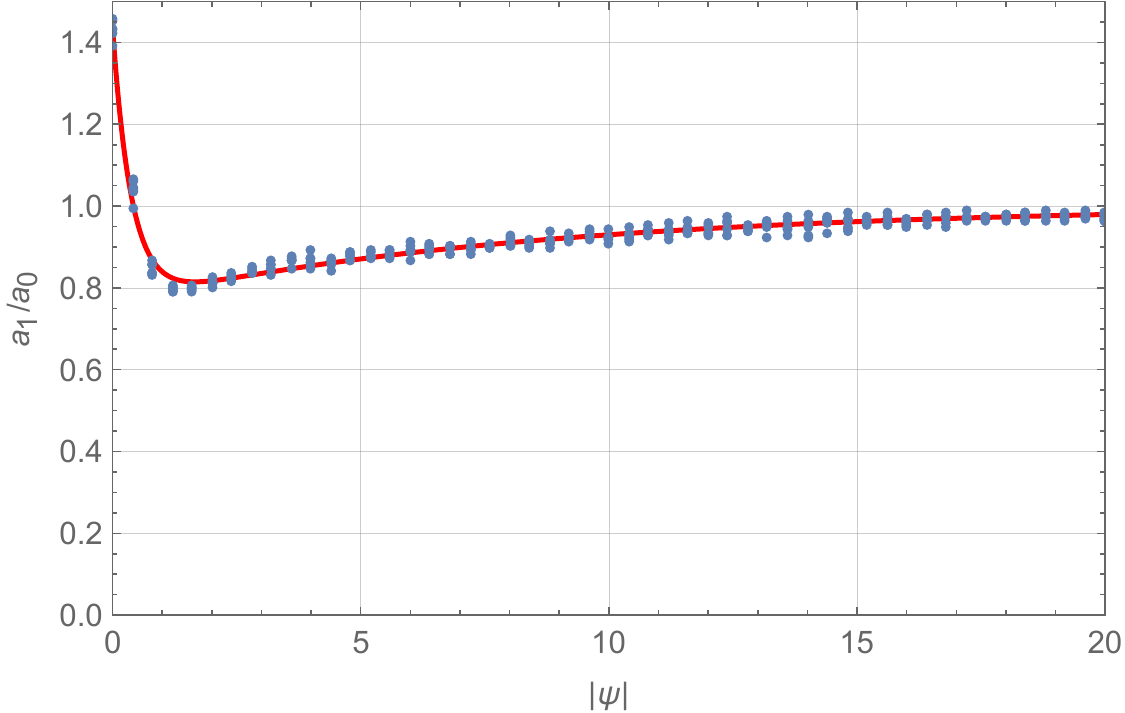}
\caption{Ratio $a_1/a_0$ of the best-fit parameters in the Ansatz~\eqref{bckp} for the bi-cubic K\"ahler potential, as a function of $|\psi|$ (blue points). The red curve is a graph of the function $f$ in Eq.~\eqref{bcf}, the best fit of a sum of two exponentials to the data.}\label{fig:bcbovera}
\end{center}
\end{figure}

The best-fit values for the coefficients $a_0$, $a_1$ and $c$ in the Ansatz~\eqref{bckp} as a function of $|\psi|$ are shown in Fig.~\ref{fig:bcabc} on the right. Evidently, the value of $c$ is $\psi$-independent to a good approximation and its mean value is $c\simeq 1/\pi$, as expected. As in our analysis of the quintic, the scatter in the data points for $a_0$ and $a_1$ in Fig.~\ref{fig:bcabc} on the right can be reduced by removing the ambiguity due to K\"ahler transformations and by plotting $a_1/a_0$ instead. This has been done in Fig.~\ref{fig:bcbovera}.
As in the quintic case the result strongly suggests that $a_1/a_0$ is a function of only the modulus $|\psi|$. It turns out that its functional behavior can be well described by a sum of two exponentials,   
\begin{equation}\label{bcf}
    f(|\psi|)=\frac{a_1}{a_0}\simeq 1+0.675 e^{-2.959 |\psi|}-0.238 e^{-0.122|\psi|}\, ,
\end{equation}
and the graph of this function is shown in Fig.~\ref{fig:bcbovera} (red curve). Hence, the expression for the K\"ahler potential can be written as
\begin{equation}\label{bckp1}
    K=\frac{t}{\pi}\ln\left(I_0+f(|\psi|)I_1\right)
\end{equation}
with the invariants $I_0$, $I_1$ from Eq.~\eqref{bcinv} and the function $f$ from Eq.~\eqref{bcf}.

Note that the blue points in Fig.~\ref{fig:bcbovera} indicate the asymptotic behavior $a_1/a_0\rightarrow 1$ for $|\psi|\rightarrow\infty$. The functional form of $f$ in Eq.~\eqref{bcf} has thus been chosen  to reflect this feature. Assuming this is indeed the correct asymptotic form, the K\"ahler potential for large $|\psi|$ becomes
\begin{equation}
    K\simeq \frac{t}{\pi}\ln\left(I_0+I_1\right)=\frac{t}{\pi}\ln\left(|x_0|^2+|x_1|^2+|x_2|^2\right)+
    \frac{t}{\pi}\ln\left(|y_0|^2+|y_1|^2+|y_2|^2\right)\; ,
\end{equation}
that is, a sum of the Fubini-Study K\"ahler potentials of the two $\mathbb{P}^2$ ambient space factors.

\section{Conclusion}\label{sec:conclusion}
In this paper, we have found analytic K\"ahler potentials for approximately Ricci-flat metrics on certain simple Calabi-Yau (CY) manifolds. Specifically, we have considered the one-parameter family of Dwork quintic CY hyper-surfaces in $\mathbb{P}^4$ and an analogous one-parameter family of bi-cubic CY hyper-surfaces in $\mathbb{P}^2\times\mathbb{P}^2$.

Our method is conceptually simple. Using previous result~\cite{Larfors:2021pbb,Larfors:2022nep} and the {\sf cymetric} package we have first calculated the (approximate) Ricci-flat K\"ahler potential numerically via machine learning methods, for a number of values of the single complex structure parameter $\psi$. These numerical results have then been compared with Donaldson's Ansatz for the K\"ahler potential for relatively low-degree underlying line bundles. Given the large symmetry groups of our manifolds only a few free parameters remain in this Ansatz. For each value of $\psi$ we have determined these parameters by a fit to the numerical data, thereby finding a K\"ahler potential close to the `optimal' one (rather than the balanced one computed by Donaldson's algorithm). The quality of the fits is quite convincing, typically reproducing the data within a few percent. This indicates that the K\"ahler potentials obtained provide a good approximation to the Ricci-flat ones. Moreover, after dealing with a subtlety related to K\"ahler transformations, we have been able to extract the $\psi$-dependence of these K\"ahler potentials. Perhaps somewhat unexpectedly, they only depend on the modulus, $|\psi|$, and appear to be independent of the phase of $\psi$, within the accuracy of our computation.

It is worth emphasising that the discrete symmetries $G$ of the CY manifolds $X$ we have considered considerably simplify the discussion. For a freely-acting $G$, the unique Ricci-flat metric on the smooth quotient $X/G$, once pulled back to $X$, results in a Ricci-flat metric, which, by construction, respects the symmetry $G$. In this case, the Ricci-flat K\"ahler potential must be $G$-invariant. This argument does not directly apply to the non freely-acting symmetries $G$ we have been using but it is natural to expect a $G$-invariant K\"ahler potential even in this case. In fact, the success of our numerical calculations supports this expectation. 

The analysis of the quintic is based on the line bundle $L={\cal O}_X(1)$ and we have discussed the cases of $L^k$, with $k=1,2,3$. For $k=1$ the optimal K\"ahler potential is simply the Fubini-Study K\"ahler potential on $\mathbb{P}^4$, restricted to the quintic. The results for the approximately Ricci-flat K\"ahler potentials for $k=2,3$ are given in Eqs.~\eqref{kpq2} and \eqref{kpq3}, respectively. For the one-parameter family of bi-cubics, we have considered the line bundle $L={\cal O}_X(1,1)$ for $k=1$ and the approximately Ricci-flat K\"ahler potential is given in Eq.~\eqref{bckp1}.

While we have been able to determine the K\"ahler potential's dependence on the complex structure parameter $\psi$ explicitly, matters are not quite so straightforward for the K\"ahler parameters. For manifolds with a single K\"ahler parameter $t$, such as the quintic CY, the K\"ahler potential (and the associated metric) scales with this parameter. 
Therefore, in such cases, the K\"ahler class dependence can easily be incorporated, as has been done in Eqs.~\eqref{kpq2} and \eqref{kpq3}. On the other hand for CY manifolds $X$ with $h^{1,1}(X)>1$, such as the bi-cubic CY with $h^{1,1}(X)=2$, the situation is more complicated. Donaldson's Ansatz based on sections of a line bundle $L\rightarrow X$ is appropriate for a K\"ahler class $[J]$ proportional to $c_1(L)$. A change in K\"ahler class requires a different line bundle and leads to a completely different Ansatz for the K\"ahler potential. It is, therefore, difficult to see how K\"ahler moduli dependence for cases with $h^{1,1}(X)>1$ can be incorporated in this approach. For our example based on the bi-cubic CY we have bypassed this difficulty by focusing on the line with $t_1=t_2$ in K\"ahler moduli space.

The K\"ahler potentials in Eqs.~\eqref{kpq2}, \eqref{kpq3} and \eqref{bckp1} are rather simple when expressed in terms of the projective ambient space coordinates. However, if we want to calculate the associated K\"ahler metric we have to remember to restrict first to a patch in the ambient space and then to the CY hypersurface, using the defining equation. This complicates matters considerably. It remains to be seen, and is the subject of ongoing work, if our results for approximate Ricci-flat K\"ahler potentials can facilitate approximate analytic calculation of metric-dependent quantities, despite these difficulties.

\section*{Acknowledgements}
SJL thanks the University of Oxford, where part of this work was completed, for hospitality. The work of SJL was supported in part by the Yonsei University Research Fund 2025-22-0134. AL thanks the Institute for Basic Science (IBS) in Daejeon, where part of this work was done, for hospitality. AL is supported in part by the STFC grant ST/X000761/1.

\end{document}